# Towards Automatic Prediction of Outcome in Treatment of Cerebral Aneurysms

Ashutosh Jadhav, Ph.D[1,*], Satyananda Kashyap, Ph.D[1,*], Hakan Bulu, Ph.D[1], Ronak Dholakia, Ph.D[2], Amon Y. Liu, MD[3], Tanveer Syeda-Mahmood, Ph.D[1], William R. Patterson, Ph.D, Hussain Rangwala, Ph.D[2], Mehdi Moradi, Ph.D[1]
[1]IBM Research - Almaden, San Jose, CA, USA; [2]MicroVention, Inc., Aliso Viejo, CA; [3]AYL Consulting LLC, Redwood City, CA, USA
*Equal contribution

**Abstract:**
*Intrasaccular flow disruptors treat cerebral aneurysms by diverting the blood flow from the aneurysm sac. Residual flow into the sac after the intervention is a failure that could be due to the use of an undersized device, or to vascular anatomy and clinical condition of the patient. We report a machine learning model based on over 100 clinical and imaging features that predict the outcome of wide-neck bifurcation aneurysm treatment with an intrasaccular embolization device. We combine clinical features with a diverse set of common and novel imaging measurements within a random forest model. We also develop neural network segmentation algorithms in 2D and 3D to contour the sac in angiographic images and automatically calculate the imaging features. These deliver 90% overlap with manual contouring in 2D and 83% in 3D. Our predictive model classifies complete vs. partial occlusion outcomes with an accuracy of 75.31%, and weighted F1-score of 0.74.*

## Introduction

A cerebral aneurysm (also known as a brain aneurysm) is a weak or thin spot on an artery in the brain that balloons or bulges out and fills with blood. The bulging aneurysm can rupture or cause mass effect on the nerves or brain tissue[1]. Different sources report rates of 3-5% of the population with asymptomatic intracranial aneurysms[2]. It is a primary cause for more than 80% of nontraumatic life-threatening subarachnoid hemorrhages[3]. A ruptured aneurysm can cause serious health problems such as hemorrhagic stroke, brain damage, coma, and death. A major advancement in endovascular aneurysm treatment came with the development of intracranial flow-diverting stents (FDS)[4]. The first product of this type was approved by the US Food and Drug Administration (FDA) in 2011[5]. Flow diverters are finely woven mesh stents that divert the flow from the aneurysm sac. The Woven EndoBridge (WEB) embolization system is approved by the FDA in late 2018[6]. The WEB device is an intrasaccular braided implant that is placed within the aneurysm sac. It can be used on ruptured and unruptured aneurysms[7].

The correct sizing of the aneurysm sac, using intra-operative digital subtraction angiography is critical in achieving complete occlusion. A clinician measures the height, dome width, and the neck size of the aneurysm sac on both lateral and anterior-posterior (AP) views on DSA images to determine the diameter/height combination of the implant to be used (Figure 1). The Raymond–Roy Occlusion Classification is a widely accepted aneurysm occlusion classification system that consists of Class I as "Complete Occlusion", Class II as "Residual neck", and Class III as "Residual aneurysm"[8]. These outcomes are typically observed a year after the procedure. In case of the WEB implant, the rate of complete occlusion is reported to be around 55%[7]. Apart from implant size, there are indications that the diameters and relative positions of parent and daughter vessels in a bifurcation aneurysm could also impact the aneurysm rupture due to their effect on the flow dynamics of the blood into the sac region[8–10]. Also, history of certain neurological disorders, gender, and smoking history are associated with an increased incidence of aneurysms[11]. While there are multiple retrospective studies to evaluate the clinical outcome of

WEB implants[12,13], automatic prediction of outcome with this kind of implant is not explored. A few new recent efforts address the problem of outcome prediction in aneurysm treatment with stents and other kinds of devices[14,15]. Our work combines multiple groups of clinical and imaging features, reports the use of automatically generated 3D measurements, and focuses on intrasaccular implants.

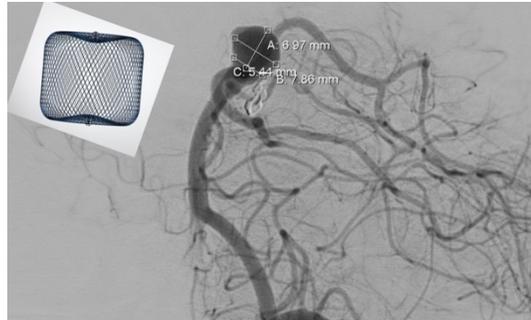

**Figure 1:** A sample 2D DSA image and WEB device (the size of the device is not proportional to the DSA image).

*A predictive model based on recent advances in artificial intelligence (AI) that can determine the treatment outcome for a patient, given that patient's clinical background and imaging studies, can help with the choice of treatment and implant type.* To that end, this work introduces two contributions: 1) First, we built a machine learning model using clinical features and imaging measurements to predict the outcome of WEB treatment for a patient. We conducted a comprehensive set of experiments with different types of machine learning classification algorithms and various combinations of over 100 clinical and imaging features. Our final Random Forest model predicts WEB implant outcome in cerebral aneurysms with an accuracy of 75.3%, weighted F1 score of 0.74 and a sensitivity of 92%. 2) Second, we built two deep neural network segmentation models to automatically segment the sac and measure the relevant features, in 2D and 3D DSA images. This is needed since some of the features included in our model are image-based measurements of the aneurysm sac such as volume and surface area. We show that the automatically calculated features are accurate based on a comparison with clinician measurements. This contribution removes the need for manual measurements to enable running the proposed model on a patient's data.

**Materials And Methods**

**Dataset:** The data used in this study was collected during the WEB-IT clinical trial[7]. The dataset consists of 81 cases across 30 different US hospitals that had obtained patient consent for research use. The WEB implant outcome after 12 months is observed on angiographic images. The distribution of these cases across three Raymond–Roy occlusion classes was as follows: "Complete Occlusion": 49 cases, "Residual Aneurysm": 5 cases, and "Residual Neck": 27 cases. Due to the limited number of cases, we have combined both adverse classes "Residual Aneurysm" and "Residual Neck" into one class as "Partial Occlusion (32 cases)". The dataset consisted of 21 male patients with average age of 60.7 years and 60 female patients with average age of 58.6 years. The distribution of the patient population (Male: M, Female: F) across the two classes were - "Complete Occlusion" (M:14, F:35) and "Partial Occlusion" (M:7, F:25).

The clinical data provided structured information about the patient's clinical profile, aneurysm information, and other data as shown in (**Table 1**). The "Pre-existing conditions" category had data about patients' existing clinical conditions. These conditions were grouped by 14 different body systems such as Cardiovascular and circulatory, Psychological/Psychiatric, Neurological, Musculoskeletal, and Gastrointestinal. For example, under the "Cardiovascular and circulatory" body system the possible health conditions were "Hypertension", "Coronary Artery Disease", "Valve Disease/Dysfunction", "Hypotension", "Arrhythmia", "Myocardial Infarction", "Angina", and "Heart Failure". The data also included 1D measurements of aneurysm (height, width, and dome in AP and lateral view) from DSA images.

| No | Category | Features |
|---|---|---|
| 1 | Demographic | Age, gender, height, weight, race |
| 2 | Aneurysm information | Location (Anterior communicating artery complex, Basilarex, Internal carotid artery terminus, Middle cerebral artery bifurcation), side (right, left, midline), type (ruptured, unruptured), the unruptured aneurysm was detected by (incidentally, symptomatically), Hunt and Hess Grade, NIHSS Score, mRS Score |
| 3 | Dimensions of aneurysm | AP and Lateral view (height, weight, neck) |
| 4 | Allergies and drugs | Known allergies and medication information |
| 5 | Pre-existing conditions | Smoking history, substance abuse, affected body systems (neurological, psychological/ psychiatric, cardiovascular and circulatory, endocrine, metabolic, musculoskeletal, eyes/ears/nose/throat/head/neck, respiratory, gastrointestinal; genitourinary, hematological/lymphatic, dermatological) |

**Table 1**: List of clinical categories and their features for the clinical data.

**Imaging Data Annotation:** An experienced neuro-interventional radiologist performed DSA image annotations using OsiriX DICOM Viewer (Pro). The dataset comprised multiple series of 2D and 3D angiographic data. The first step was to curate data, modify DICOM headers to separate series, and identify relevant series. The clinician identified the working AP and the lateral projections, where available, adding up to 144 2D images from the 81 cases, with some missing the lateral view. Then, the following annotations were performed:

1) Annotations on 2D DSA images for dimensions and contour of the aneurysm sac: As shown in (Figure 2-A), these include the height, the width, and the dome length of the aneurysm region in two orthogonal DSA projections (AP and lateral), to deliver six dimensions. For the predictive analysis experiments, we have calculated sum of 3 dimensions (height, width, and dome) for AP and lateral view. We also use the area of the aneurysm sac in these DSA images as a measurement. To calculate this area, the clinician marks the contour. These contours are also used to train a neural network to automate the segmentation and measurement process.

2) Annotations of vessel geometry, size, and angle at bifurcation: The diameters of the daughter branches and of the parent vessel, and the angles between each daughter vessel (dominant and non-dominant) and the parent vessel, were measured based on the markings by the clinician on images as shown in (Figure 2-B). We calculated angle between parent vessel and left daughter (vessel - left angle) as well as angle between parent vessel and right daughter (vessel - right angle). Further, we computed normalized variants of these angles such as vessel - normalized left angle = (vessel - left angle)/180. We calculated diameters of parent vessel, left daughter branch, right daughter branch and larger daughter branch, and the ratio of these measurements.

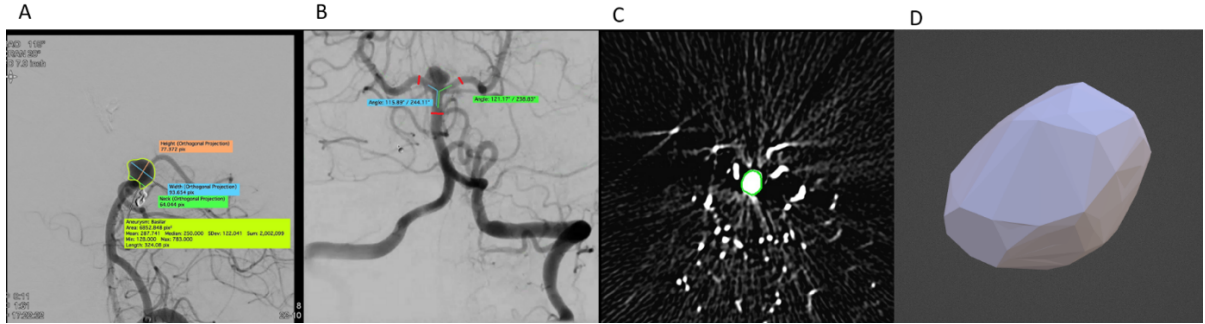

**Figure 2:** Annotation process and examples. (A) Annotations of dimension and contour in the working projection 2D DSA images. (B) Vessel annotations (diameter and angles) in 2D DSA images. (C) Sample of contour annotation in axial slice stack for 3D sac reconstruction. (D) The 3D mesh model of the sac reconstructed from segmented contours

<u>3) Annotations of vessel geometry, size, and angle at bifurcation:</u> The diameters of the daughter branches and of the parent vessel, and the angles between each daughter vessel (dominant and non-dominant) and the parent vessel, were measured based on the markings by the clinician on images as shown in (Figure 2-B). We calculated angle between parent vessel and left daughter (vessel - left angle) as well as angle between parent vessel and right daughter (vessel - right angle). Further, we computed normalized variants of these angles such as vessel - normalized left angle = (vessel - left angle)/180. We calculated diameters of parent vessel, left daughter branch, right daughter branch and larger daughter branch, and the ratio of these measurements.

<u>4) 3D annotations on axial slice stack:</u> In 41 of the available cases, a stack of reconstructed axial slices was available from the imaging study. The clinician contoured the perimeter of the sac in all images in the stack where it was visible. This was used to reconstruct a volumetric mesh model of the sac and to estimate the volume and surface area of the sac (Figure 2-C and D). Note that this data was also used to train another separate neural network model to segment the 3D sac for automatic calculation of volumetric measurements.

**Derived Image-based Features:** For some cases, lateral view measurements were missing in the data. To address this missing measurement problem, we computed "aggregated" measurements from AP and lateral view as follows:

$$Agg = \begin{cases} \dfrac{AP + Lat}{2}, & \text{if both exist} \\ AP, & \text{Lateral missing} \\ Lat, & \text{AP missing} \end{cases}$$

For 3D features, surface area and the volume of the aneurysm sac were calculated from the 3D mesh models. From these, we also calculated Non-sphericity index (NSI) and Isoperimetric ratio (IPR) as follows: *Non-sphericity index = $1 - (18\pi)^{1/3}(V^{2/3}/S)$* and *Isoperimetric ratio = $S/(V^{2/3})$*, with *V* and *S* referring to the aneurysm volume and area, respectively.

**Automatic Calculation of Imaging Feature using Deep Neural Net Segmentation Models:**

The manual annotation of images for the purpose of calculating the imaging features is not always practical in clinical settings. As a result, we explored automating the calculation of 2D and 3D measurements of area and volume. The primary step to be automated was the delineation of the aneurysm boundary by an expert neuro-interventional radiologist. This is a time consuming and expensive endeavor which needs to be performed in a time critical situation of implant selection. Automatic or semi-automated segmentation methods can reduce the burden on the doctors and potentially help reduce measurement time and implant selection. To this end, we explored using state of the art deep learning-based algorithms to automatically segment the aneurysm regions from the images and use them as features in the predictive modelling. We propose two deep learning-based algorithms for 2D (DSA) and 3D (axial stacks) images respectively to accurately contour the aneurysm sac volume.

**2D DSA Segmentation Algorithm:** We use a UNet[16] like architecture in which we replace the encoder arm of the UNet with a pre-trained state of the art EfficientNet[17] architecture. The advantage of this setup is we are able to leverage a network trained on several millions of images as a rich feature generator that can be fine-tuned for our specific task of segmentation. We used a semi-automated pre-processing pipeline wherein a rectangular region with heuristic padding is created around the aneurysm based on minimal input from the radiologist. The deep learning models were implemented in Python's PyTorch package with the network architecture implementation based on[18]. A k-fold cross validation *(k=15, Train: 134 ± 1.72, Test: 9.6 ± 1.72 images)* was performed on image sizes 1024×1024×3, with batches of 16 images for 20 epochs with an initial learning rate of $1e^{-3}$, reduced learning rate on plateau scheduler. Weak augmentation of flipping, rotation, Gaussian blurring was done to increase the variety of training inputs.

**3D Axial Stack Segmentation Algorithm:** We used an extended version of the 3D UNet called the AH-Net[19] to segment the 3D data. The network leverages pre-trained ResNet50[20] architecture to generate encoder features. The feature decoder uses anisotropic 3D convolutions thereby learning the contextual information across the 2D slice-by-slice features generated from the encoder. This allows the network to leverage well trained 2D feature generators even for complex 3D tasks. The pre-processing pipeline used a semi-automated method of extracting a heuristically determined cube around the aneurysm based on minimal input from a radiologist. The pipeline was implemented in PyTorch with the AHNet architecture implementation from Monai package[21]. Weak augmentations of randomaffine, randomanisotropy[22] were used. A k-fold cross validation *(k=15, train: 38.27 ± 0.44, valid: 2.73 ± 0.44)* was performed on image sizes 60×60×60, with batches of 12 for 30 epochs with a learning rate of $1e^{-3}$. For both deep learning methods Dice coefficient[23] loss was used to optimize the network.

**Feature Selection:** We used two feature selection approaches – statistical co-occurrence and information gain. The co-occurrence-based approach is used to select relevant health conditions from the 67 possible health conditions in 14 different affected body systems (see **Table 1**). We selected health conditions as determinants of outcome based on their *differential prevalence* in the "Complete Occlusion" and "Partial Occlusion" classes. To that end, we computed the ratio of the co-occurrence of a given health condition with the total number of cases for the class. For example, for "Hypertension" the ratio of *(Hypertension/Complete Occlusion)* is *(28/49 = 0.57)* and *(Hypertension/Partial Occlusion)* is *(19/32 = 0.59)* while for the "Migraines" the ratio of *(Migraines/Complete Occlusion)* is *(28/49 = 0.57)* and *(Migraines/Partial Occlusion)* is *(14/32 = 0.44)*. We filtered-out the health conditions that have a ratio under 0.30 based on inputs from a clinician. The conditions that are remaining after this filtration are the candidate conditions for the classification task. To select discriminative conditions between the two classes from the candidate conditions, we computed an absolute difference between the prevalence ratios from two classes.

We retained only the conditions that have the difference 0.10 or above. For example, we selected "Migraines" that had absolute difference of *0.13* between the two ratios *(|0.57 − 0.44|)* while filtered out "Hypertension" that had absolute difference of *0.02* between the two ratios *(|0.57 − 0.59|)*. We use information gain as a metric to rank features[24]. Information gain metric ranks features independently for their class separability. In practice, features could have a different rank when used in a set for classification. In order to identify such feature groups, while keeping the top features (with information gain above 0.15), we also experimented with adding various combinations of other clinical and imaging features in a greedy fashion based on the weighted F1 score in 10-fold cross validation.

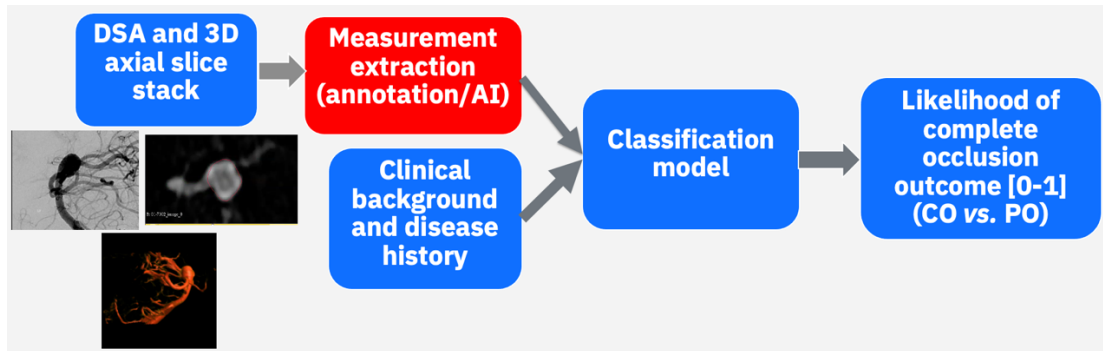

**Figure 3:** Predictive model building workflow

**Experimental Set up and Feature Sets:** We experimented with different types of classification algorithms and various combinations of the clinical and imaging features for partial versus complete occlusion classification **(Figure 3)**. We examined these five types of binary classification algorithms: random forest, multilayer perceptron (MLP) neural network, logistic regression, naive Bayes, and support vector machine (SVM) as implemented in Weka Java library[25,26] which handles missing values using C4.5 algorithm[27]. Finally, the optimal algorithm is selected based on 10-fold cross-validation and performance measures (F1- score, sensitivity, specificity, and ROC).

During the process of building the predictive model, we have developed multiple models considering different combinations of the clinical and imaging features. "Feature set A" has all the clinical and pre-operative imaging features and it is used to develop a baseline model. "Feature set B" contains no imaging features and it is a subset of clinical features that are selected to maximize class separability as described above. "Feature set C" contains the selected clinical features same as "Feature set B" but adds to those the selected imaging features from 2D DSA images as calculated directly from clinical annotations. "Feature set D" contains the same features as "Feature set C" with the addition of the aneurysm sac volume-related measurements extracted from 3D image annotations. To evaluate the effects of automatic measurement, we also experimented with "Feature set E" which replaces the contour measurement from 2D and volume, surface and IPR from 3D features with their equivalents calculated from automatically delineated contours that were obtained using the neural nets. Note that clinical features are same for "Feature set B", "Feature set C", "Feature set D", and "Feature set E". Finally, to study the effect of dropping clinical features, we used "Feature set F" which contains only imaging features from "Feature set D" and "Feature set G" which only uses imaging features from "E". "Feature set E", and "Feature set G" uses automatically computed 3D imaging using deep neural net segmentation model.

## Results

**Automatic Segmentation Results:** The results from the segmentation models are shown in **(Table 2)**. For the 2D DSA we compared our results with the Baseline UNet, UNet with EfficientNet-B0 with and without pre-training. Our results show that the UNet with a pre-trained encoder resulted in the best dice score of 0.9 on average. For the 3D segmentation we compare AHNet with a baseline UNet and a slice-by-slice implementation of the 2D EfficientNet-B1 method. While the slice-by-slice method provided reasonable accuracy, the use of 3D context in the AHNet helped improve the results even further to 0.83. Sample qualitative results of the two deep learning methods are shown in **(Figure 4).**

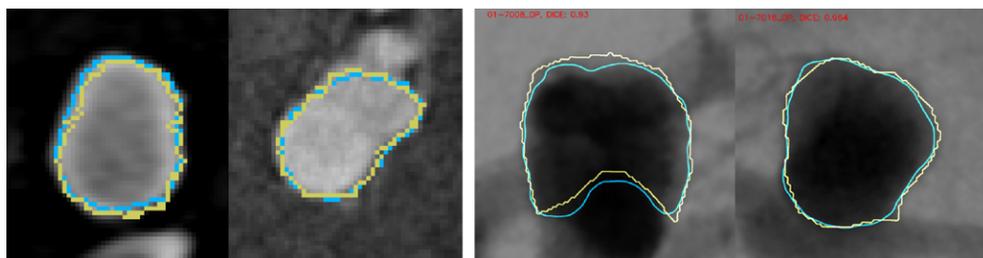

**Figure 4:** Sample results of proposed 3D (left) and 2D (right) methodology for segmentation of detected aneurysm sac. The ground truth contour (yellow) and the contour produced by the proposed automated (blue) are shown.

| Method | Task | Dice (95% CI) |
|---|---|---|
| **2D model (n=144)** | | |
| Baseline: UNet | 2D segmentation | 0.628 (0.592-0.663) |
| UNet, EfficientNet-B0 encoder (w/o pre-train) | 2D segmentation | 0.864 (0.849-0.879) |
| UNet, EfficientNet-B0 encoder | 2D segmentation | **0.908 (0.899-0.918)** |
| **3D model (n=41)** | | |
| Baseline 3D: UNet | 3D segmentation | 0.675 (0.620-0.730) |
| UNet, EfficientNet-B1 encoder | 3D segmentation, slice-by-slice | 0.810 (0.775-0.844) |
| AHNet 3D, ResNet50 encoder | 3D segmentation | **0.836 (0.813-0.859)** |

**Table 2**: The average and standard deviation of the Dice coefficient for our 2D segmentation model, proposed methods and baseline (first 3 rows) and for our 3D segmentation model, proposed methods and baseline (last three rows). CI: Confidence Interval.

**Outcome Prediction Results (Table 3):** We found that the random forest is consistently the best performing classifier for our dataset and classification problem. We noticed a significant increase in performance from "Feature set A" to "Feature set B" (p-value for weighted F1 score: 0.03). The performance further improved with the addition of annotation-based 2D AP and lateral view measurements, and vessel geometry-related measurements ("Feature set C") and aneurysm 3D sac volume-related measurements ("Feature set D"). Our best performing random forest model using "Feature set D" has an accuracy of 75.3% (as compared to 58.02% baseline "Feature set A"), 0.74 weighted F1 scores (as compared to 0.52 baseline "Feature set A" model). The random forest model outperformed all other classification approaches. The average weighted F1 score with 10-Fold cross-validation on "Feature set D" for the other approaches was: MLP (0.61), logistic regression (0.64), naïve Bayes (0.64), and SVM (0.51). The weighted F1 score dropped merely by 1% with the replacement of annotation-based aneurysm sac volume-related measurements with automatically calculated measurements ("Feature set D" *vs.* "Feature set E"). Finally, we experimented with only imaging features of feature sets "D" and "E". Interestingly we found that the F1 score obtained using imaging-only feature set with automatically calculated measurements ("Feature set G") is higher than that for the feature set with manual measurements "Feature set F") with weighted F1 scores of 0.66 and 0.59 respectively.

| Feature Set | Accuracy (95% CI) | Specificity (95% CI) | Sensitivity (95% CI) | Weighted F1 Score (95% CI) | ROC (95% CI) | F1 (CO) (95% CI) | F1 (PO) (95% CI) |
|---|---|---|---|---|---|---|---|
| A | 58.0% (47.3%-68.8%) | 15.6% (7.7% - 23.5%) | 85.7% (78.1% - 93.3%) | 0.52 (0.41 - 0.63) | 0.50 (0.39-0.61) | 0.71 (0.61-0.81) | 0.23 (0.14 - 0.32) |
| B | 66.7% (56.4% - 76.9%) | 43.8% (33.0% - 54.6%) | 81.6% (73.2% - 90.1%) | 0.65 (0.55–0.75) | 0.58 (0.47 - 0.69) | 0.75 (0.66-0.84) | 0.51 (0.40-0.62) |
| C | 72.8% (63.2% - 82.5%) | 50.0% (39.1% - 60.9%) | 87.8% (80.6% - 94.9%) | 0.72 (0.62 – 82) | 0.66 (0.56-0.76) | 0.80 (0.71-0.89) | 0.59 (0.48-0.70) |
| D | **75.3% (65.9% - 84.7%)** | **50.0% (39.1% - 60.9%)** | **91.8% (85.9% - 97.8%)** | **0.74 (0.64–0.84)** | **0.71 (0.61–0.81)** | **0.82 (0.74-0.90)** | **0.62 (0.51-0.73)** |
| E | 74.1% (64.5% - 83.6%) | 53.1% (42.3% - 64.0%) | 87.8% (80.6% - 94.9%) | 0.73 (0.63–0.83) | 0.68 (0.58-0.78) | 0.80 (0.71-0.89) | 0.62 (0.51-0.73) |
| F | 60.5% (49.8% - 71.1%) | 37.5% (27.0% - 48.0%) | 75.5% (66.1% - 84.9%) | 0.59 (0.48–0.70) | 0.56 (0.45-0.67) | 0.70 (0.60-0.80) | 0.43 (0.32–0.54) |
| G | 67.9% (57.7% - 78.1%) | 43.8% (33.0% - 54.6%) | 83.7% (75.6% - 91.7%) | 0.66 (0.56–0.76) | 0.63 (0.53-0.74) | 0.76 (0.67-0.85) | 0.52 (0.41–0.63) |

**Table 3**: Feature set performances with Random Forest, 10-Fold cross validation. CO: Complete Occlusion. PO: Partial Occlusion.

**Discussions**

In this work, we built a model using clinical and imaging features to predict the outcome of the treatment of wide-neck bifurcation aneurysm with an intrasaccular embolization device. Some of the clinical features identified as correlates of bad outcome here have not been reported in previous work. These include neurological conditions history, psychological/psychiatric conditions. This finding deserves further study with a larger sample size. Statistical comparison of the prediction results obtained from different feature sets provides insights about the importance and effectiveness of features. The comparisons discussed below are all based on weighted F1-score and a T-test on iterations of the random forest classifier. The baseline "Feature set A" is consistently the least predictive feature set ($p-value < 0.001$ in performance comparison with feature sets "C", "D" and "E", which is significant even after accounting for multiple comparisons effect, and $p-value < 0.05$ in comparison with "B", significant only without counting for multiple comparison effect). It is notable that the imaging features alone ("Feature set G") outperform the clinical features alone ("Feature set B"). We also demonstrated the effectiveness of the neural network segmentation methods to outline the aneurysm sac in 2D and 3D DSA images to produce image-based features automatically. The difference between F1-score of image-based measurements obtained by clinical annotations ("Feature set D") versus the set that uses the automatically calculated version of the same measurements (E) is *not* significant ($p-value = 0.45$). Results from "Feature set E" and "G" are important as they highlight how deep learning based medical imaging analysis can generate measurements that are comparable to an experienced neuro-interventional radiologist's annotation. The utilization of automated measurements facilitates the use of our decision support system without the need for expensive manual annotation. To further understand the possible causes of unfavorable outcome, we computed the difference between the measured volume of the sac versus the volume of the fully expanded WEB device that was implanted. For the "Complete Occlusion" group, the average difference is ($0.004 \pm 0.052$ $cm^3$) while for the "Partial Occlusion" group it is ($0.031 \pm 0.055$ $cm^3$). In other words, the under sizing in "Partial Occlusion" outcome group is evident in our dataset. This emphasized the importance of correct sizing in the outcome of the procedure.

**Limitation:** Our work is limited by the sample size and continued work with larger dataset is required to accumulate further insights. Specially the size of the population with partial occlusion is small (32 out of 81). As a result, our classifier tends to perform better in predicting "Complete Occlusion" as opposed to predicting "Partial Occlusion" as evident from sensitivity and specificity data in **(Table 3)**. It is promising that the rate of prediction improvement obtained by our methodology in comparison to the baseline method of "Feature set A" is larger for the "Partial Occlusion" class (improvement from 0.23 to 0.62 in F1-score) compared to "Complete Occlusion" class (improvement from 0.71 to 0.82 in F1-score).

## Conclusion

Cerebral aneurysms remain among the most acute health conditions of the modern life. Decision support software systems that help choose the most promising treatment at the rapid pace required to minimize the impacts of the disease on the patient are critical. Our work provides early insights and technologies to build such AI driven software. Our study shows that a combination of clinical and imaging features to predict WEB implant outcome in cerebral aneurysms is more efficient than using either clinical or imaging features. We showed that deep learning can be used for segmenting the aneurysm sac and producing accurate measurements for use in predictive modeling. The future work in this study includes expanding the sample size and evaluating similar approaches in prediction of outcome with other implant devices.